\documentclass[11pt]{article}
\usepackage{amsmath}
\begin{document}

\thispagestyle{empty}

\begin{flushright} LPTENS-06/06 \end{flushright}

\vskip 0.5cm

\begin{center}{\LARGE {CAN WE PREDICT THE VALUE
\vskip 0.4cm
OF THE HIGGS MASS?}}

\vskip1cm

A pedagogical exercise

\vskip1cm

{\bf{J. Iliopoulos}}
\vskip0.2cm

Laboratoire de Physique Th\'eorique CNRS-ENS\\ 
24 rue Lhomond, F-75231 Paris Cedex 05, France\\
ilio@lpt.ens.fr
\vskip1.0cm

{\bf ABSTRACT}

\end{center}

\vskip 0.5cm

In the framework of the Standard Model the mass of the physical Higgs boson is an arbitrary parameter. In this note we examine whether it is possible to determine the ratio of $m_H /M$, where $M$ denotes any other mass in the theory, such as the $W$ or the $Z$-boson mass. We show that no such relation can be stable under renormalisation.

\bigskip

\newpage

The full electroweak theory known as ``The Standard Model'' contains a large number of arbitrary parameters which must be determined by experiment. Therefore, any attempt to reduce that number by finding some relation among some of these parameters will be most welcome, provided this relation will be stable against higher order corrections and will not be spoiled by arbitrary counterterms. In this note we want to examine under which conditions this may be possible.

Let us first formulate a general problem: We consider the interaction among a set of fields $\phi_i(x)$, $i=1,...,N$, where $i$ denotes the fields as well as their space-time and/or internal symmetry indices. We consider the most general renormalisable Lagrangian among the fields $\phi_i(x)$ consistent with whatever symmetries we have chosen. The requirement of renormalisability restricts the Lagrangian density to be a polynomial in the fields and their first derivatives of degree not higher than four. The full Lagrangian density is of the form:

\begin{equation}
\label{genL}
{\cal {L}}={\cal {L}}(\phi_i, \partial_{\mu}\phi_i;m_j,g_k)
\end{equation}

We see that ${\cal {L}}$ depends on a set of arbitrary parameters which we have chosen to write as $m_j$, $j=1,...,N_m$ which are parameters with the dimensions of a mass, and $g_k$, $k=1,...,N_g$, which are dimensionless and are usually called coupling constants. It is clear that such a distinction is largely a matter of convention and we can often trade a pair of mass parameters with a mass and a coupling constant. By assumption, ${\cal {L}}$ is the most general renormalisable Lagrangian among the given set of fields, this means that the standard renormalisation procedure will not force us to introduce new terms with new arbitrary constants to any order of perturbation theory. In this sense the set of parameters $m_j$ and $g_k$ is complete. The question we want to ask in this note is whether this set is also {\it irreducible}. Roughly speaking we want to know whether we can describe the interaction among the same set of fields  $\phi_i$ using a smaller number of parameters. More precisely, we want to investigate whether we can impose conditions of the form

\begin{equation}
\label{cond}
C(m_j,g_k)=0
\end{equation}
which will remain stable under renormalisation. In practice, this often implies the existence of a renormalisation scheme which explicitly respects the condition (\ref{cond}) and does not require the introduction of a specific counterterm for it. Here however, neither the knowledge nor the construction of such a scheme will be necessary for our argument. 

In order to illustrate the argument let us show a well-known field
theory example in
which such a relation is indeed possible \cite{phiex}: It consists of two
scalar fields $\phi_1(x)$ and $\phi_2(x)$ with quartic interactions.
 To keep the example as
simple as possible, let us impose the discrete symmetries
$\phi_i(x)\rightarrow -\phi_i(x)$, $i=1,2$ and $\phi_1(x)\leftrightarrow
\phi_2(x)$. The most general renormalisable Lagrangian for this model is:

\begin{equation}
\label{mod}
{\cal{L}}  = \frac{1}{2}(\partial_{\mu}\phi_1)^2+\frac{1}{2}(\partial_{\mu}\phi_2)^2-\frac{1}{2}m^2(\phi_1^2+\phi_2^2)
 - \frac{\lambda}{4!}(\phi_1^4+\phi_2^4)-
\frac{2g}{4!}\phi_1^2 \phi_2^2
\end{equation}

The model contains two arbitrary coupling constants, $\lambda$ and $g$,  and we
can ask the question whether there exists a relation among them

\begin{equation}
\label{simplerel}
\eta=\frac{g}{\lambda}=C
\end{equation}
with a constant $C$ which is stable under
renormalisation. By inspection, we see immediately that here the answer is
yes. In fact $C=0$ is an obvious such relation, since, at this point, the
model describes a system of two  uncorrelated scalar fields. Are there any
other, less trivial, relations? The renormalisation group offers a systematic
way to address this question.

The coupling constants $\lambda$ and $g$ and their ratio $\eta$, depend on the
renormalisation scheme. This dependence is perturbative, which means that the
values at any scheme can be computed as a formal power series in the values at
any other. Let us consider two such schemes with coupling constants
$\lambda_{(1)}$, $g_{(1)}$ and  $\lambda_{(2)}$, $g_{(2)}$. We have:

\begin{equation}
\begin{split}
\label{rengr}
\lambda_{(1)}= &
\lambda_{(2)}+a_1\lambda_{(2)}^2+a_2g_{(2)}^2+a_3\lambda_{(2)}g_{(2)}+...\\
g_{(1)}= & g_{(2)}+b_1\lambda_{(2)}^2+b_2g_{(2)}^2+b_3\lambda_{(2)}g_{(2)}+...
\end{split}
\end{equation}
where the $a$'s and the $b$'s are calculable numbers which relate the two schemes, (for example, they may depend on the ratio $\mu_1/\mu_2$ of the subtraction points in the two schemes) and the dots stand for higher loop terms. 
The important, although trivial, observation, is that at lowest order we must
always have $\lambda_{(1)}=\lambda_{(2)}$ and $g_{(1)}=g_{(2)}$ since
there is no renormalisation in the classical approximation. Furthermore,
(\ref{rengr}) exhausts all possible choices in the sense that, given a set of
coupling constants $\lambda$ and $g$ obtained in a scheme (1), any other possible
set will be given in terms of a set of numbers $a$ and $b$.  The scheme dependence (\ref{rengr}) is governed by the renormalisation group. A straightforward calculation gives for the $\beta$-functions associated with the coupling constants $\lambda$ and $g$  at the one loop level:

\begin{equation}
\label{beta}
\begin{split} 
16 \pi^2 \beta_{\lambda}= & 3\lambda^2+\frac{1}{3}g^2\\
16 \pi^2 \beta_{g}= & \frac{4}{3}g^2+2\lambda g
\end{split}
\end{equation}

A relation invariant under renormalisation is a fixed point of the renormalisation group, {\it i.e.} a zero of the corresponding $\beta$-function. For example, we find immediately the trivial fixed point $g=0$, a zero of $\beta_{g}$, which means that if we start with no $g$ term in the Lagrangian, no such term will be generated by renormalisation.  If we are interested in the behaviour of the ratio  $\eta=g/\lambda$ we compute the $\beta$-function

\begin{equation}
\label{eta} 
\beta_{\eta}=  \frac{\lambda \beta_{g}-g \beta_{\lambda}}{\lambda^2}
 =  \frac{-1}{48\pi^2}\eta \lambda  (\eta^2-4\eta+3) 
\end{equation}

Eq. (\ref{eta}) shows that we obtain three fixed points, namely $\eta$=0, 1 and
3. The first is the trivial one we saw before. The second ($\eta$=1), gives a
more symmetric theory in which the discrete symmetries of (\ref{mod}) are
promoted into a continuous $O(2)$ group with $\phi_1$ and $\phi_2$ becoming
the two components of an $O(2)$ vector. Finally the third fixed point,
$\eta$=3, through the transformation $\psi_1=\phi_1+\phi_2$ and
$\psi_2=\phi_1-\phi_2$ reduces again to a system of two uncoupled fields with
interaction $\psi_1^4+\psi_2^4$. Furthermore, eq. (\ref{eta}) shows that these
are the only possible values of the ratio of the coupling constants which are
respected by renormalisation.

For a theory with more than two
coupling constants the solution requires the study of all possible relations which are admissible in perturbation theory, a task whose weight increases very fast with the number
of coupling constants. For every relation, we compute the corresponding $\beta$-function and look for possible fixed points. Although an exhaustive study may be quite tedious, the answer is always unambiguous. 
The moral of the story is that
the reducibility of a model, in other words the possibility of imposing
renormalisation stable relations among its parameters, can be studied by
studying the zeros of the $\beta$-functions. 
One may ask the question: how
reliable is such a computation? Answer: As reliable as perturbation theory can
possibly be. The reason is the well known result that the first 
non-vanishing terms in the expansion of the $\beta$-functions are independent
of the renormalisation scheme one uses to compute them. A simple proof of this
result is presented at the end of this note. 
\vskip 0.5cm

After these preliminaries, let us go back to the Standard electroweak Model. 
In order to simplify the discussion, let us first restrict ourselves to the
bosonic sector of the model, leaving out all quarks and leptons.  In this
simple sector the model is determined by four independent parameters. In the
unbroken phase they can be chosen to be $g_1$ and $g_2$, the gauge coupling
constants for the groups $U(1)$ and $SU(2)$  respectively, $\lambda$, the
strength of the scalar Higgs field quartic self-coupling and $\mu^2$, the
scalar field mass term which is the only dimensionfull parameter of the
model. If $\mu^2<0$ spontaneous symmetry breaking occurs and three out of the
four gauge bosons become massive leaving one physical neutral, spin zero
particle, the famous {\it Higgs boson}. In this phase it is convenient to
trade $\mu$ for $v$, the vacuum expectation value of the Higgs field. In the
tree approximation the two are related by $v^2=-\mu^2/\lambda$. This relation
may receive corrections at higher orders in perturbation theory, but they will
be finite and calculable. 

The mass spectrum of this model in the classical approximation is given by:\\
1) A massless neutral vector boson which can be identified to the photon.\\
2) A charged pair of spin one bosons, $W^{\pm}$ with mass $m_W=vg_2/2$. They are
the intermediate vector bosons of the charged current weak interactions, such
as $\beta$-decay\\
3) A massive neutral vector boson $Z^0$ with mass
$m_Z=v \sqrt{g_2^2+g_1^2}/2=m_W/cos\theta_W$, with $tan\theta_W=g_1/g_2$.\\
4) A massive neutral spin-zero particle $\phi^0$, the Higgs particle, with
mass $m_{\phi}=v\sqrt{2\lambda}$. 

We see that the model predicts definite relations among the various masses
with all ratios expressed in terms of the coupling constants. The
renormalisation properties of the theory guarantee that these relations will
receive only finite and calculable corrections at any given order of
perturbation theory. All but one of these parameters have already been
over-determined by experiment, for example, $m_W$, $m_Z$ and $\theta_W$ are
measured independently. Such measurements offer a splendid confirmation of the
Standard Model. The only parameter which has not yet been directly measured is the
Higgs boson mass, or, alternatively, the coupling constant $\lambda$. In the
Standard Model it is an arbitrary parameter and any relation between
$m_{\phi}$ and any other mass should be sought at physics beyond the Standard
Model. In this note I want to argue that, unless one enlarges the particle
content of the model, no such relation will be stable against
renormalisation. 

\vskip 0.5cm

Let us assume that such relation exists and determines the ratio, for example,
$m_{\phi}/m_Z=C$, with $C$ some constant. At the classical level such relation is indeed obtained if one formulates the model in a suitably chosen space with non-commutative geometry which allows for a unified picture of both Higgs and gauge fields. \cite{ncgeom}. In the tree approximation we have:

\begin{equation}
\label{relation}
C=\frac{m_Z}{m_{\phi}}=\frac{\sqrt{g_1^2+g_2^2}}{\sqrt{8\lambda}}
\end{equation}

So, the question is: is there any renormalisation scheme, no matter how
complicated in practice, in which the relation (\ref{relation}) does not
receive an infinite counterterm? As we pointed out before, such a relation
will correspond to a zero of the $\beta$-function for the combination of the
coupling constants which appears at the r.h.s. of (\ref{relation}). Similar investigations, searching for fixed points in the framework of the Standard Model, or extensions of it, have been performed already \cite{other}. Here we focus only on the mass ratios of the Higgs scalar and the gauge bosons.   The
$\beta$-functions of the Standard Model, with or without fermions, have been
computed at one and two loops \cite{betafun}. Without fermions we obtain:

\begin{equation}
\label{smbeta1}
\begin{split}
16\pi^2\beta_{g_1} & =g_1^3\frac{1}{10}\\
16\pi^2\beta_{g_2} & =-g_2^3\frac{43}{6}\\
16\pi^2\beta_{\lambda} &
=12\lambda^2-\frac{9}{5}g_1^2\lambda-9g_2^2\lambda+\frac{27}{100}g_1^4+\frac{9}{10}g_1^2g_2^2+\frac{9}{4}g_2^4
\end{split}
\end{equation}

With three coupling constants we can form the two ratios:

\begin{equation}
\label{smeta1}
\eta_1=\frac{g_1^2}{\lambda} ~~~;~~~ \eta_2=\frac{g_2^2}{\lambda}~~~;~~~ z=\eta_1+\eta_2~~~;~~~\rho=\frac{\eta_1}{\eta_2}
\end{equation}

Notice that $\lambda$ must be positive, otherwise the classical Higgs
potential is unbounded from below. This implies that both $\eta$'s are
positive. The corresponding $\beta$-functions are given by:

\begin{equation}
\label{smeta2}
\begin{split}
\beta_{\eta_1} & =\frac{1}{\lambda^2}(2g_1\lambda
\beta_{g_1}-g_1^2\beta_{\lambda})=\\
 & =\frac{\lambda}{16 \pi^2}\left (2\eta_1^2-12 \eta_1+9 \eta_1 \eta_2
-\frac{27}{100}\eta_1^3-\frac{9}{10}\eta_1^2 \eta_2-\frac{9}{4}\eta_1
\eta_2^2 \right )\\
\beta_{\eta_2} & =\frac{1}{\lambda^2}(2g_2\lambda
\beta_{g_2}-g_2^2\beta_{\lambda})=\\
 & =-\frac{\lambda}{16 \pi^2}\left (\frac{16}{3}\eta_2^2+12\eta_2-\frac{9}{5}\eta_1 \eta_2+\frac{27}{100}\eta_1^2 \eta_2+\frac{9}{10}\eta_1
\eta_2^2+\frac{9}{4}\eta_2^3 \right)
\end{split}
\end{equation}

Using (\ref{smeta2}) we can investigate any ratio of the Higgs and the gauge
boson masses. For the combination (\ref{relation}) we obtain:

\begin{equation}
\label{smz1}
\begin{split}
\beta_z & =\beta_{\eta_1}+\beta_{\eta_2}=\\
 & =\frac{-\lambda w}{16\pi^2 \rho z}\left [ \left
 (\frac{27}{100}\rho^2+\frac{9}{10}\rho+\frac{9}{4}\right )z^2+\left
 (2\rho^2+\frac{54}{5}\rho -\frac{16}{3}\right )z-12(\rho +1)^2\right ]
\end{split}
\end{equation}

It is easy to check that the quadratic form in the r.h.s. of (\ref{smz1})
never vanishes for real and positive $z$ and $\rho$. This implies that the
relation (\ref{relation}) will be violated in one loop, no matter which
renormalisation scheme one is using. 

Similarly, we can check whether the ratio $m_W/m_{\phi}$ can provide a stable
fixed point by looking for possible zeros of the $\beta$-function of $\eta_2$,
eq. (\ref{smeta2}). The result is again negative. In fact, since we set up all this formalism, we can address a more general question: Is
there any generalised mass ratio of the form
$z_{\theta}=cos{\theta}\eta_1+sin{\theta}\eta_2$ which gives a stable
relation? For this to happen the corresponding $\beta$-function $\beta_{z_{\theta}}=cos{\theta}\beta_{\eta_1}+sin{\theta}\beta_{\eta_2}$ must vanish at a point $z_{\theta}=C$, with $C$ some, possibly $\theta$-dependent, constant. It is easy again to check that there is no such fixed point.

The introduction of fermions does not help in producing fixed points. Their presence has two effects: (i) They change the running of the coupling constants by adding new contributions to the $\beta$-functions. For example, for three families, the coefficient of $\beta_{g_2}$ becomes -31/6 with similar changes for the other $\beta$-functions. (ii) They introduce new terms in $\beta_{\lambda}$. They are due to the Yukawa couplings between the Higgs scalar and the fermions which are responsible for the fermion masses. In practice only the top-quark coupling is numerically important because of its large value. In turn, the quark diagrams bring the strong interactions through the one gluon exchange contributions. As a result $\beta_{\lambda}$ now depends on many more coupling constants, namely, $\lambda$, $g_1$ and $g_2$, but also  $h_Y$, the entire set of the fermion Yukawa couplings. They have their own evolution equations which depend also on $g_3$, the QCD coupling \cite{betafun}. In two loops the evolution equations of all coupling constants are coupled. Since the argument \cite{ncgeom} was independent of $g_3$ and $h_Y$, the constant $C$ of the relation (\ref{relation}) should also be independent of them. It follows that, even if the theory had a fixed point in the absence of fermions, their presence would have made it unstable. 

The conclusion is that the set of parameters of the Standard Model appears to be irreducible. Any relation among them will unavoidably be violated by quantum corrections. This does not mean that it is impossible to predict the mass of the Higgs boson. It only means that the origin of such a relation should come from physics beyond the Standard Model in which the latter is embedded in a larger scheme with tighter structure and, most important, richer particle content.

\vskip 1cm

\noindent {\it Proof of the scheme independence of the first two terms in the
  expansion of the $\beta$-function:} Let us assume a theory with only one
  coupling constant.  In one renormalisation scheme we have the function
  $\beta(\lambda )$ with an expansion

\begin{equation}
\label{vita}
\beta(\lambda )=b_0 \lambda^2+b_1 \lambda^3+...
\end{equation}

By changing the renormalisation scheme we obtain a new coupling constant
$\lambda'$ and a new $\beta$-function $\beta'(\lambda')$. The two are related
by:

\begin{equation}
\label{newcc}
\lambda'=F(\lambda)=\lambda+f_1 \lambda^2+O(\lambda^3)
\end{equation}

\begin{equation}
\label{newvita}
\beta'(\lambda')=\mu \frac{\partial}{\partial \mu} \lambda'=\frac{\partial
  F}{\partial \lambda} \beta(\lambda )
\end{equation}

Notice again the fact that the first term in the expansion of $F(\lambda)$ in
(\ref{newcc}) is $\lambda$. It is this property that makes $F$ a new acceptable
coupling constant.

We expand now both sides of (\ref{newvita}) and find:

\begin{equation}
\label{exp}
\begin{split}
\beta'(\lambda')=(1+2f_1 \lambda + O(\lambda^2))(b_0 \lambda^2+b_1
\lambda^3+O(\lambda^4))\\
=b_0 \lambda^2+(b_1+2f_1 b_0)\lambda^3+O(\lambda^4)=b_0 \lambda'^2+b_1
\lambda'^3+O(\lambda'^4)
\end{split}
\end{equation}
where we have used the inverse relation implied by (\ref{newcc}), namely
$\lambda=\lambda'-f_1 \lambda'^2+O(\lambda'^3)$. We have thus established the
universality of $b_0$ and $b_1$. The generalisation to several coupling constants is
  straightforward. The result is again that the first non-vanishing term is scheme independent. Notice that the same proof shows that the
first non-vanishing term in the expansion of the anomalous dimension of the
field $\gamma (\lambda)$ is also scheme independent. 

A final remark: The relation (\ref{newvita}) shows that the existence of a
zero of $\beta$ is a universal property, although the particular value of the
coupling constant for which the zero occurs is scheme dependent. 

\vskip 1cm

\noindent {\Large \bf Acknowledgements}

\vskip 0.5cm

The author wishes to thank Drs. M. Dubois-Violette and F. Zwirner for discussions.

\end{document}